\documentclass[twocolumn,showpacs,preprintnumbers,amsmath,amssymb]{revtex4}

\usepackage{amsmath,amssymb}

\usepackage{color}
\usepackage[dvips]{graphicx}
\usepackage{subfigure}

\newcommand{\lsim}{\mathrel{\mathop{\kern 0pt \rlap
  {\raise.2ex\hbox{$<$}}}
  \lower.9ex\hbox{\kern-.190em $\sim$}}}
\newcommand{\gsim}{\mathrel{\mathop{\kern 0pt \rlap
  {\raise.2ex\hbox{$>$}}}
  \lower.9ex\hbox{\kern-.190em $\sim$}}}
\newcommand{\sigmav}{\langle \sigma_{\rm ann} v \rangle}

\newcommand{\beq}{\begin{equation}}
\newcommand{\eeq}{\end{equation}}
\newcommand{\bea}{\begin{eqnarray}}
\newcommand{\ena}{\end{eqnarray}}
\newcommand{\etal}{{\it et al.}}

\renewcommand{\prd}[3]{Phys.\ Rev.\ D\ {\bf #1}, #3 (#2)}

\begin{document}

\preprint{DFTT 08/2005}
\preprint{LAPTH--1104/05}

\title{Antiproton fluxes from light neutralinos}



%
\author{A. Bottino}
\affiliation{Dipartimento di Fisica Teorica, Universit\`a di Torino \\
Istituto Nazionale di Fisica Nucleare, via P. Giuria 1, I--10125 Torino, Italy}
\author{F. Donato}
\affiliation{Dipartimento di Fisica Teorica, Universit\`a di Torino \\
Istituto Nazionale di Fisica Nucleare, via P. Giuria 1, I--10125 Torino, Italy}

\author{N. Fornengo}
\affiliation{Dipartimento di Fisica Teorica, Universit\`a di Torino \\
Istituto Nazionale di Fisica Nucleare, via P. Giuria 1, I--10125 Torino, Italy}

\author{P. Salati}
\affiliation{
Laboratoire d'Annecy-le-Vieux de Physique Th\'eorique LAPTH\\
CNRS-SPM and Universit\'e de Savoie\\
9, Chemin de Bellevue, B.P.110 74941 Annecy-le-Vieux, France}
\date{\today}

\begin{abstract}
We analyze how the measurements of the low-energy spectrum of
cosmic antiprotons can provide information on relic neutralinos.
The analysis is focused on the light neutralinos which
emerge in supersymmetric schemes where gaugino-mass unification
is not assumed. We determine which ranges of the
astrophysical parameters already imply stringent constraints 
on the supersymmetric configurations and those ranges 
which make the antiproton
flux sensitive to the primary component generated
by the neutralino self-annihilation. Our results are derived
from some general properties of the antiproton flux
proved to be valid for a generic cold dark matter candidate.

\end{abstract}


\maketitle


\section{Introduction}
\label{sect:intro}
     In supersymmetric schemes, where gaugino-mass unification is not assumed,
the lower bound on the neutralino mass is determined by the upper limit on the
contribution of cold dark matter (CDM) to the cosmological density parameter,
$(\Omega_{\rm CDM})_{\rm max}h^2$. 
Using the value $(\Omega_{\rm CDM})_{\rm max} h^2 = 0.13$,
derived from results of Refs. \cite{wmap,sloan}, one obtains the lower bound
$m_{\chi} \gsim 7$ GeV \cite{lowneu,lowmass}. This is at variance with the more
commonly employed lower limit of about 50 GeV, derived from the experimental
(LEP) lower bound on the chargino mass. The rich phenomenology related to
possible light neutralinos, with masses within the range 7 GeV $\leq m_{\chi}
\leq $ 50 GeV, has been discussed in Refs. \cite{lowneu,lowmass,lowdir,lowind}.

In particular, in Ref. \cite{lowind} it was scrutinized the capability of indirect measurements of
WIMPs (Weakly Interacting Massive Particles)
to detect light neutralinos and it was concluded that measurements of cosmic antiprotons
 represent the most promising mean of indirect exploration. The analysis of the antiproton
signal was based on a study of the antiproton flux at one single representative value of the
antiproton kinetic energy ($T_{\bar{p}}$ = 0.23 GeV). It was proved that, for a wide range of the
astrophysical parameters, the antiproton signal due to neutralino 
pair-annihilation is within the
level of detectability at small values of $m_{\chi}$.

In the present paper we extend the previous investigation of the antiproton signal by analyzing the
detailed features of the expected theoretical spectra, and discuss how our results can be employed
to determine the presence of an actual primary signal or, at least, to derive  significant
constraints on supersymmetric configurations at small $m_{\chi}$. Our analysis is carried out in the
perspective of a significant breakthrough in the determination of some relevant astrophysical
parameters and in view of a sizeable improvement in the measurement of the cosmic antiprotons
spectrum, as expected in forthcoming experiments in space. 

This paper is organized as follows. In Sect. \ref{sect:generic} we analyze
some properties of primary cosmic antiproton fluxes due to self-annihilation of
a generic candidate of cold dark matter. These features are derived in the
standard scheme usually employed to describe the decoupling
of cold particles from the primordial plasma. An upper bound is
obtained for the antiproton flux. Then, in Sect. \ref{sect:neu}, a full
evaluation of the antiproton spectrum is derived in the case of relic 
neutralinos. Results and perspectives are presented in Sect. \ref{sect:results}.

\section{A few properties of the self-annihilation cross-section for
 a generic WIMP}
\label{sect:generic}

Cosmic primary antiprotons  can originate from the hadronization of quarks and
gluons produced in WIMP self-annihilation processes
(we consider here selfconjugate WIMPs) \cite{bulk,pbar0,Bottino_Salati,bergstrom,pbar_susy}.
Once antiprotons are produced in the
dark halo, they diffuse and propagate throughout the Galaxy.  The  propagated
antiproton differential
flux at a generic point of coordinates $r, z$ in the Galactic rest frame ($r$ is the
radial distance from the Galactic center in the Galactic plane and $z$ is the vertical coordinate)
is

\begin{equation}
 \Phi_{\bar{p}} (r,z,E) = \frac{v_{\bar{p}}}{4\,\pi} \; \Upsilon \;
\frac{dN}{d T_{\bar{p}}}
S^{\bar{p}}_{\rm astro} (r,z,E),
\label{eq:flux}
\end{equation}

\noindent
with
\begin{equation}
\Upsilon = \frac{1}{2}  \xi^2 \, \frac{\sigmav_0}{m_\chi^2}.
\label{eq:upsilon}
\end{equation}

\noindent Notations are as follows: $v_{\bar{p}}$ is the
antiproton velocity, $\sigmav_0$ is the average, over the Galactic
velocity distribution, of the WIMP annihilation cross-section
multiplied by the relative velocity.
 $ S^{\bar{p}}_{\rm astro}$ is a function which
takes into account all the effects of propagation in the Galaxy and
includes $\rho^2(r,z)$, $\rho (r,z)$ being the galactic dark matter 
distribution. Here we take:

\begin{equation}
\rho(r,z) = \rho_l \frac{a^2 + R_\odot^2}{a^2+r^2+z^2},
\end{equation}

\noindent 
where $\rho_l$ is the total local dark matter density
set at the value of 0.3 GeV/cm$^3$, $a$ is a core parameter, $a$=3.5 kpc 
and $R_\odot$=8 kpc.
In the following we do not include any clumpiness effect. 
$\xi$ represents the fractional local density $\rho_\chi$ of our
generic WIMP as compared to  $\rho_l$, {\it i.e.} $\xi = \rho_\chi/\rho_l$.
By applying the usual rescaling procedure \cite{gst},  one has $\xi =
\min[1,\Omega_{\chi} h^2/(\Omega_{\rm CDM}h^2)_{\rm min}]$. 
From the analyses of Refs. \cite{wmap,sloan} one derives that at
$2\sigma$ level the cosmologically interesting region for cold dark matter
is: $0.095 \leq \Omega_{\rm CDM} h^2 \leq 0.13$ (in what follows this will be
denoted as the WMAP range for CDM abundance). Thus, for 
$(\Omega_{\rm CDM}h^2)_{\rm min}$ we use here the value 
$(\Omega_{\rm CDM}h^2)_{\rm min} = 0.095$.

 $d N_{\bar p}/ d T_{\bar p}$
is the differential antiproton spectrum per annihilation event:
\begin{equation}
\frac{d N_{\bar p}}{d T_{\bar p}} =
\sum_F BR(\chi \chi \rightarrow {\bar p} + X) \frac{dN_{\bar p}^{F}}{dT_{\bar p}},
\label{eq:spectrum}
\end{equation}

\noindent
where $F$ denotes the different annihilation final states, $BR(\chi \chi
\rightarrow {\bar p} + X)$ the branching ratios and $dN_{\bar p}^{F}/dT_{\bar
p}$ stands for the antiproton energy spectra in the $F$ channel. 

Throughout this paper we will be interested in the antiproton differential flux
at Earth ($r = R_\odot$ , $z$ = 0) as a function of the antiproton kinetic energy; 
this flux will be simply denoted  as $\Phi_{\bar{p}}(T_{\bar{p}})$.

We turn now to a discussion about some general properties of the quantities
entering the flux factor $\Upsilon$.
To make the discussion more transparent, we consider a scenario where the standard expansion
 in S and P waves for the thermally averaged product of
the annihilation cross-section times the relative velocity of the self-interacting particles

\begin{equation}
<\sigma_{\rm ann} v> \; \simeq \; \tilde{a} + \tilde{b} \; \frac{1}{x},
\label{sigmav}
\end{equation}
\noindent
holds ($x$ is defined as $x=m_\chi/T$, $T$ being the temperature). 

For relic particles  in the Galactic halo $x \sim 10^6$, then, usually, a good
approximation is:
\begin{equation}
<\sigma_{\rm ann} v>_0 \; \simeq \; \tilde{a}
\label{sigmav0}
\end{equation}
We recall that $<\sigma_{\rm ann} v>$ enters also in the relic abundance:
\begin{equation}
\Omega_{\chi} h^2 = \frac{x_f}{{g_{\star}(x_f)}^{1/2}} \frac{3.3 \times
10^{-38} \; {\rm cm}^2}{\widetilde{<\sigma_{\rm ann} v>}},
\label{omega}
\end{equation}
where $\widetilde{<\sigma_{\rm ann} v>} \equiv x_f {\langle \sigma_{\rm
ann} \; v\rangle_{\rm int}}$, ${\langle \sigma_{\rm ann} \;
v\rangle_{\rm int}}$ being the integral
of $<\sigma_{\rm ann} v>$ from the present temperature up to the freeze-out temperature $T_f$;
$x_f$ is defined as $x_f \equiv m_{\chi}/T_f$ and
${g_{\star}(x_f)}$ denotes the relativistic degrees of freedom of the
thermodynamic
bath at $x_f$.  Using the expansion of Eq. (\ref{sigmav}), one obtains:
$\widetilde{\langle \sigma_{\rm ann} \; v\rangle}\simeq \tilde{a} +
1/(2 x_f) \tilde{b}$.
Since $x_f \simeq 20$, also the P-wave contribution $\tilde{b}$ has to be retained in
this case. In the following, we however specifically assume that
$\tilde{a} \geq 1/(2 x_f) |\tilde{b}|$.

\subsection{Lower bound on $<\sigma_{\rm ann} v>_0$}

    The cosmological upper bound
$\Omega_{\chi} h^2 \leq (\Omega_{\rm CDM} h^2)_{\rm max}$
implies that  $\widetilde{\langle \sigma_{\rm ann} \; v\rangle}$ 
is limited from below. For a cold relic with a mass in the range
10 GeV $\lsim m_{\chi} \lsim $ 1 TeV, one has
${g_{\star}(x_f \simeq  20)} \sim 90$, so that
$x_f / {g_{\star}(x_f)}^{1/2} \sim 2.2$ (with variations of order 10 \%).
Then, from Eq. (\ref{omega})
$\widetilde{\langle \sigma_{\rm ann} \; v\rangle}
\gsim 7.3 \times 10^{-38} {\rm cm}^2/(\Omega_{\rm CDM} h^2)_{\rm max}$. Using the value
$(\Omega_{\rm CDM} h^2)_{\rm max} = 0.13$, we obtain
$\widetilde{\langle \sigma_{\rm ann} \; v\rangle} \gsim 5.6 \times 10^{-37}$ cm$^2$.
If $\tilde{a} \geq 1/(2 x_f) |\tilde{b}|$, this implies

\begin{equation}
<\sigma_{\rm ann} v>_0 \; \gsim \;  3 \times 10^{-37} {\rm cm}^2.
\label{sigmav0b}
\end{equation}

\subsection{Upper bound on $\xi^2  <\sigma_{\rm ann} v>_0$}
\label{sect:upper}

Often it turns out that $<\sigma_{\rm ann} v>_0$ may be orders of magnitude
larger than the lower limit of Eq. (\ref{sigmav0b}). However, this fact does
not automatically imply very large values for the antiproton flux, since the
relevant quantity which enters in the antiproton flux is not simply
$<\sigma_{\rm ann} v>_0$, but instead  $\xi^2  <\sigma_{\rm ann} v>_0$, through
the factor $\Upsilon$.  Indeed, the quantity $\xi^2  <\sigma_{\rm ann} v>_0$
coincides with $<\sigma_{\rm ann} v>_0$  when $\Omega_{\chi} h^2\geq
(\Omega_{\rm CDM} h^2)_{\rm min}$, but it is proportional to $<\sigma_{\rm ann}
v>_0/{\widetilde{\langle \sigma_{\rm ann} \; v\rangle}}^2$ when $\Omega_{\chi}
h^2< (\Omega_{\rm CDM} h^2)_{\rm min}$; then it has a maximum at $\Omega_{\chi}
h^2= (\Omega_{\rm CDM} h^2)_{\rm min}$ \cite{bffms,lathuile}. Let us call
$\eta$ the set of parameters of the particle-physics model which describes our
generic cold relic. The property, that we have just discussed, states that the
maximum of the quantity $\xi^2  <\sigma_{\rm ann} v>_0$:
\begin{equation}
(\xi^2  <\sigma_{\rm ann} v>_0)_{\rm max} = <\sigma_{\rm ann} v>_0|_{\eta = \eta'}
\end{equation}
\noindent
occurs when the model parameters $\eta$ have values $\eta'$, such that
$(\Omega_{\chi} h^2)_{\eta = \eta'} =  (\Omega_{\rm CDM} h^2)_{\rm min}$,
that is (using Eq. (\ref{omega})), when

\begin{equation}
\left(\frac{{g_{\star}(x_f)}^{1/2}}{x_f} 
{\widetilde{<\sigma_{\rm ann} v>}}\right)_{\eta = \eta'} =
\frac{3.3 \times 10^{-38} \; {\rm cm}^2}{(\Omega_{\chi} h^2)_{\rm min}}.
\label{eq:max}
\end{equation}

\noindent
Using the estimate already employed above, {\it i.e.}
$x_f / {g_{\star}(x_f)}^{1/2} \sim 2.2$, from Eq. (\ref{eq:max}) one obtains

\begin{equation}
({\widetilde{<\sigma_{\rm ann} v>}})_{\eta = \eta'} =
\frac{7.3 \times 10^{-38} \; {\rm cm}^2}{(\Omega_{\chi} h^2)_{\rm min}} \simeq
7.7 \times 10^{-37} \; {\rm cm^2},
\label{eq:max1}
\end{equation}

\noindent
where, in the last step, the value $(\Omega_{\chi} h^2)_{\rm min} = 0.095$ is used.
Thus, {\it within a factor of 2}, the maximum of $\xi^2  <\sigma_{\rm ann} v>_0$ is stable for all sets
of parameters $\eta$ which satisfy Eq. (\ref{eq:max1}) and is given by
\begin{equation}
(\xi^2  <\sigma_{\rm ann} v>_0)_{\rm max} \simeq  8 \times 10^{-37} \; {\rm cm^2}.
\label{bound}
\end{equation}
 If we take the mass $m_{\chi}$
to be one of the model parameters, this limit is
independent of $m_{\chi}$, provided the other parameters vary within ranges which allow
solutions  of Eq. (\ref{eq:max1}).

As a consequence of the previous properties, we find that
the maximum of the factor $\Upsilon$, as a function of
$m_{\chi}$,  is expected to decrease simply as 
$\Upsilon_{\rm max} \propto  m_{\chi}^{-2}$.

We note that the above properties apply also to
any other primary flux of cosmic particles due to WIMP self-interactions in
the halo, {\it e. g.} to the gamma-ray or positron flux.

\section{Primary cosmic antiproton flux from self-annihilation of light
neutralinos}
\label{sect:neu}

Now we finalize our previous considerations to the case of light relic
neutralinos.

\subsection{The Supersymmetric Model}
\label{sect:susy}

    The supersymmetric scheme employed here is an
 effective Minimal Supersymmetric extension of the Standard Model (MSSM)
 at the electroweak scale, where gaugino-mass unification is not assumed. This
 is defined in terms of a minimal number of parameters, only those
necessary to shape the essentials of the theoretical structure of MSSM
and of its particle content. The assumptions that we impose at the
electroweak scale are: a) all squark soft-mass parameters are
degenerate: $m_{\tilde q_i} \equiv m_{\tilde q}$; b) all slepton
soft-mass parameters are degenerate: $m_{\tilde l_i} \equiv m_{\tilde
l}$; c) all trilinear parameters vanish except those of the third
family, which are defined in terms of a common dimensionless parameter
$A$: $A_{\tilde b} = A_{\tilde t} \equiv A m_{\tilde q}$ and
$A_{\tilde \tau} \equiv A m_{\tilde l}$.  As a consequence, the
supersymmetric parameter space consists of the following independent
parameters: $M_2, \mu, \tan\beta, m_A, m_{\tilde q}, m_{\tilde l}, A$
and $R \equiv M_1/M_2$.  In the previous list of parameters we have
denoted by $\mu$ the Higgs mixing mass parameter, by $\tan\beta$ the
ratio of the two Higgs v.e.v.'s,  by $m_A$ the mass of the CP-odd
neutral Higgs boson, and by $M_1, M_2$ the U(1), SU(2) gaugino masses,
respectively.

In the numerical random scanning of the supersymmetric parameter
space we use the following ranges: $1 \leq \tan \beta \leq 50$,
$100~ {\rm GeV }\leq |\mu|$, $M_2 \leq 1000~{\rm GeV }$, $100~{\rm
GeV}\leq m_{\tilde q}, m_{\tilde l} \leq 1000~{\rm GeV }$, ${\rm
sign}(\mu)=-1,1$, $90~{\rm GeV }\leq m_A \leq 1000~{\rm GeV }$,
$-3 \leq A \leq 3$, in addition to the above mentioned range $0.01
\leq R \leq 0.5$. We impose the experimental constraints:
accelerators data on supersymmetric and Higgs boson searches,
measurements of the $b \rightarrow s + \gamma$ decay and of the
branching ratio $B_s \rightarrow \mu^+ \mu^-$, and measurements of the
muon anomalous magnetic moment $a_\mu \equiv (g_{\mu} - 2)/2$. For
the ranges used for these observable and other details of the
model we refer to Ref. \cite{lowind}.

\subsection{The supersymmetric flux factor $\Upsilon$}

\begin{figure}[t]
{\includegraphics[width=\columnwidth]{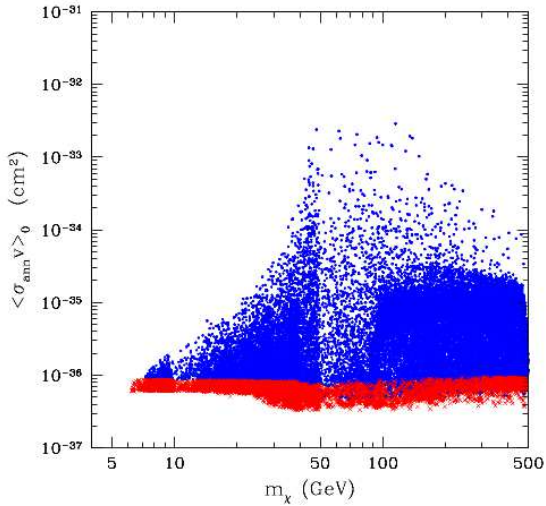}}
\caption{Scatter plot of $ <\sigma_{\rm ann} v>_0$ vs $m_{\chi}$. Red crosses
denote the supersymmetric configurations whose relic abundance is in the range 
$0.095\leq  \Omega_\chi h^2 \leq 0.13$, while blue dots denote configurations
with $\Omega_\chi h^2 \leq 0.095$.
}
\label{fig:sigmav0}
\end{figure}

\begin{figure}[t]
{\includegraphics[width=\columnwidth]{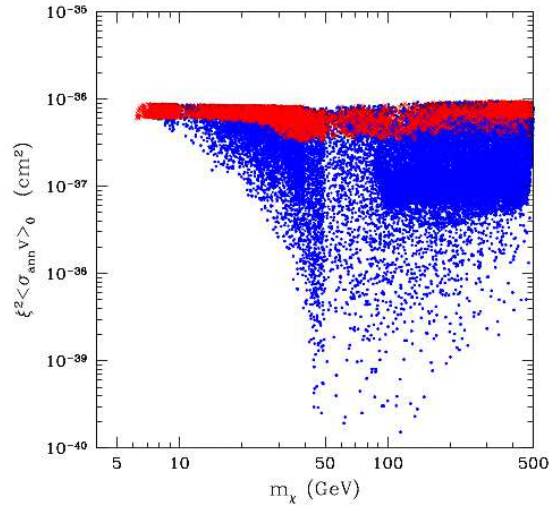}}
\caption{Scatter plot of $ \xi^2 <\sigma_{\rm ann} v>_0$ vs $m_{\chi}$. 
Notations as in Fig. \ref{fig:sigmav0}.
}
\label{fig:xi2_sigma0}
\end{figure}

\begin{figure}[t]
{\includegraphics[width=\columnwidth]{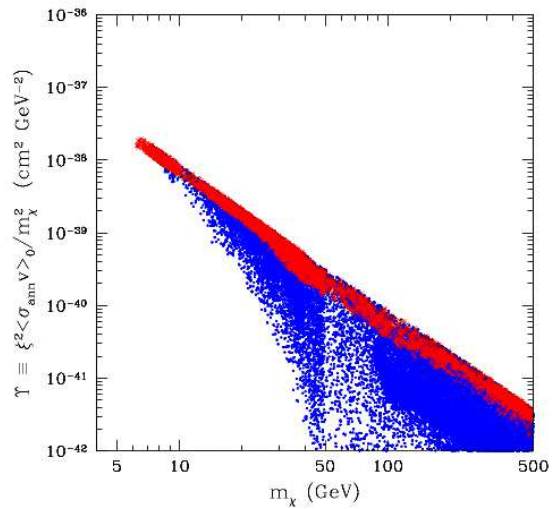}}
\caption{Scatter plot of the quantity $\Upsilon$ vs $m_{\chi}$. 
Notations as in Fig. \ref{fig:sigmav0}.
}
\label{fig:upsilon}
\end{figure}

Figs. \ref{fig:sigmav0}, \ref{fig:xi2_sigma0}, \ref{fig:upsilon} give the
scatter plots of the quantities $<\sigma_{\rm ann} v>_0$, $\xi^2<\sigma_{\rm ann} v>_0$
and $\Upsilon$ versus $m_{\chi}$, limited to supersymmetric configurations
which satisfy the approximation   $\tilde{a} \geq 1/(2 x_f) |\tilde{b}|$
employed above.  (Red) crosses denote configurations with a relic abundance in
the cosmologically relevant range $0.095 \leq \Omega_{\chi} h^2 \leq 0.13$,
(blue) dots denote susy configurations where rescaling is effective ({\it i.e.} 
neutralinos form a subdominant species of relic particles). In Fig.
\ref{fig:sigmav0} we notice the effect of the lower bound on $<\sigma_{\rm ann}
v>_0$, implied by the cosmological upper bound on $\Omega_{\rm CDM}h^2$ (see Eq.
(\ref{sigmav0b})). The rapidly rising of the scatter plot as $m_{\chi}$ reaches
the value of 45 GeV is due to the self-annihilation process through the
$Z$-boson exchange, superimposed to a similar enhancement extending to higher
values of $m_{\chi}$, originating in the self-annihilation process with the
exchange of the lightest CP-even neutral Higgs boson $h$. The upper frontier of
the scatter plot at low values of $m_{\chi}$ is determined by the experimental
lower bound on the mass of this Higgs boson.

Fig. \ref{fig:xi2_sigma0} displays the upper
bound on $\xi^2  <\sigma_{\rm ann} v>_0$, whose approximate estimate is given in Eq. (\ref{bound}).

Finally, we note that the upper frontier of the scatter plot for $\Upsilon$ in Fig.
\ref{fig:upsilon} clearly displays the simple
 behavior $\Upsilon_{\rm max} \propto  m_{\chi}^{-2}$, as derived in Sect. \ref{sect:upper}.
 The lower part of the plot is composed by
 configurations with a large rescaling in the local density.

\subsection{The differential antiproton spectrum $d N_{\bar p}/ d T_{\bar p}$}

\begin{figure*}[t]
{\includegraphics[width=\columnwidth]{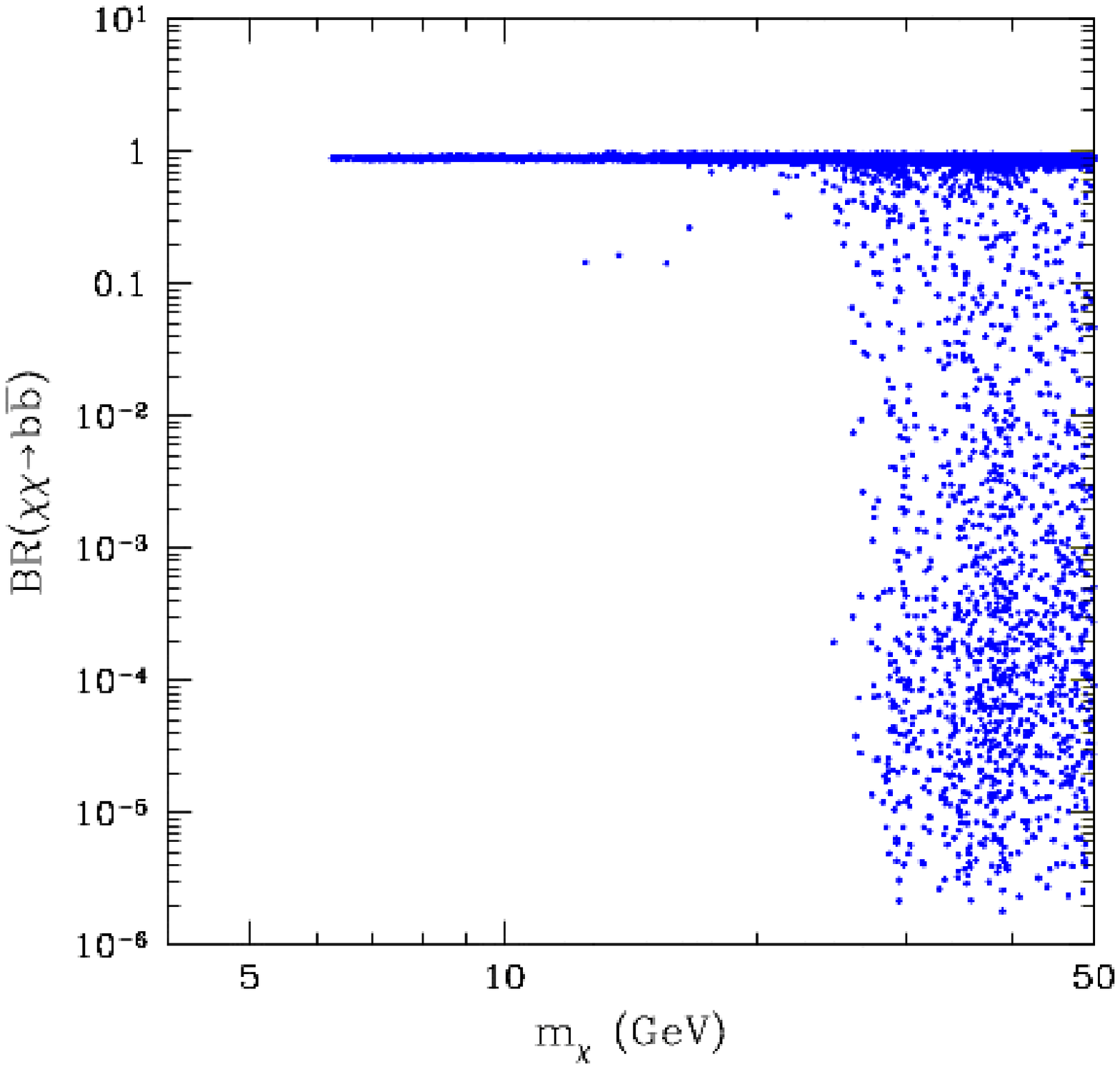}}
{\includegraphics[width=\columnwidth]{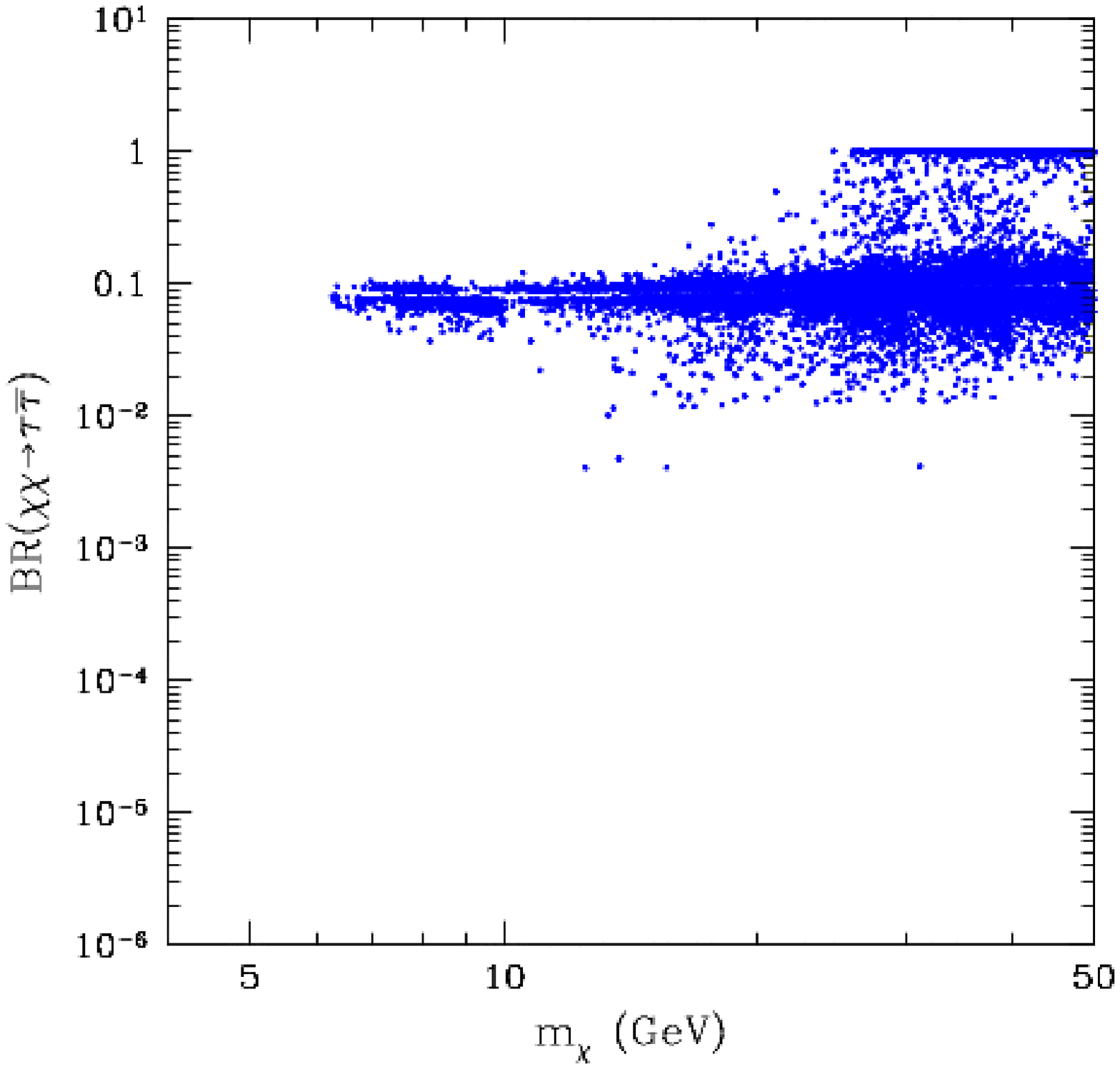}}
\caption{Scatter plots of the branching ratios for the neutralino 
self-annihilation into $b\bar{b}$ (left panel) and $\tau^+\tau^-$ vs 
$m_\chi$ (right panel). 
}
\label{fig:BR}
\end{figure*}

To evaluate $d N_{\bar p}/ d T_{\bar p}$, we follow
the treatment of Ref. \cite{lowind}. In case of neutralino masses below the
thresholds for gauge-bosons, Higgs-bosons and $t$ quark production, antiprotons
originate from the hadronization into $\bar{p}$'s of the quark and gluon pairs
produced in the neutralino self-annihilation. For light neutralinos
($m_{\chi} \leq$ 50 GeV), which
are mainly binos with a slight mixing with a higgsino component
\cite{lowmass},  the dominant final states are
the ones into $b \bar{b}$ and into $\tau^- \tau^+$,
the channel into $b \bar{b}$ being largely prominent for
$m_{\chi} \lsim$ 25 GeV.
This property is displayed in Fig. \ref{fig:BR}.
Our calculation of the energy
spectra has been performed by using a Monte Carlo simulation with the PYTHIA
package \cite{pithia}.

\begin{figure}[t]
{\includegraphics[width=\columnwidth]{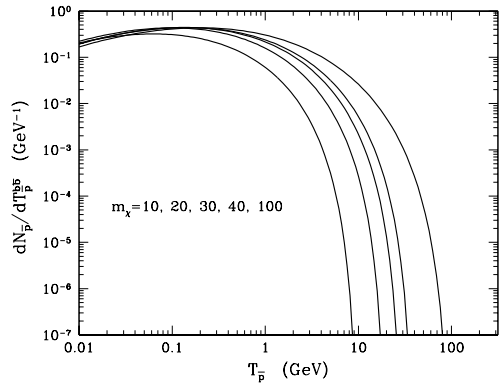}}
\caption{The differential antiproton spectrum $d N_{\bar p}/ d T_{\bar p}$
for annihilation in the $b\bar{b}$ channel, as a function 
of the antiproton kinetic energy $T_{\bar p}$. From left to 
right: $m_\chi$=10, 20, 30, 40 and 100 GeV.}
\label{fig:spectrum}
\end{figure}

For neutralino masses which kinematically allow other final states
(gauge-bosons, Higgs-bosons and $t\bar{t}$ pairs), the full decay
chain down to the production of quarks and gluons has been
evaluated analytically. The final antiproton spectrum is then
calculated from the previous results by boosting the differential
energy distribution to the rest frame of the annihilating
neutralinos.
 Details of our procedure are given in Refs. \cite{lowind,pbar_susy}.
A sample of results is shown in Fig. \ref{fig:spectrum}, where  the spectra of
antiprotons per annihilation event are shown for a sample of the neutralino
masses and self-annihilation into $b\bar{b}$. 


\subsection{Cosmic ray propagation in the Galaxy}
\label{subsect:cr_propagation}

As they propagate throughout the Galaxy, charged cosmic rays mostly
bounce on the spatial irregularities of its magnetic fields. That process
is well described by mere diffusion with the energy dependent coefficient
\beq
K \; = \; K_{0} \beta \, {\cal R}^{\delta} \;\; ,
\eeq
where ${\cal R}$ stands for the particle rigidity. The acceleration of
primary species by supernovae driven shock waves as well as their
subsequent interactions with the interstellar gas take place in a thin
galactic disk that is sandwiched above and beneath by two large diffusion
layers with thickness $L$. Because the magnetic field inhomogeneities
-- actually the Alfv\'en waves -- move with speed $V_{A}$, cosmic rays
undergo inside the disk some diffusive reacceleration that come into play
with the ionization, Coulomb and adiabatic energy losses. Particles are
finally wiped away by a vertical convective wind whose velocity is $V_{c}$.
The propagation of charged cosmic rays can be well accommodated in a
cylindrical two-zone diffusion model. The particle radial abundances
may be expanded as series of Bessel functions
$J_{0}(\alpha_{i} \, r / R_{\rm gal})$ where $\alpha_{i}$ is the i-th
zero of the function $J_{0}$ and where $R_{\rm gal} = 20$ kpc is the radius
of the propagation region. For a complete description of the semi-analytic
code on which the present analysis is based, we refer the reader
to Ref.~\cite{PaperI,PaperI_bis,revue}.
Suffice it to say that the five cosmic ray propagation parameters
mentioned above -- namely the diffusion coefficient normalization
$K_{0}$ and index $\delta$, the confinement layers thickness $L$
and the velocities $V_{c}$ and $V_{A}$ -- have been constrained
by comparing the flux predictions for various cosmic ray species with
observations. The most stringent observable is the boron to carbon ratio
B/C -- a typical secondary to primary relative abundance -- whose analysis
within our diffusion model has been detailed in Ref.~\cite{PaperI,PaperI_bis}.
The relevance of the latter has been further established by the compatibility
of the B/C results with several other observed cosmic ray
abundances \cite{PaperI_bis,pbar_sec,beta_rad}.

\begin{figure}[h!]
\vskip -2cm
{\includegraphics[width=\columnwidth]{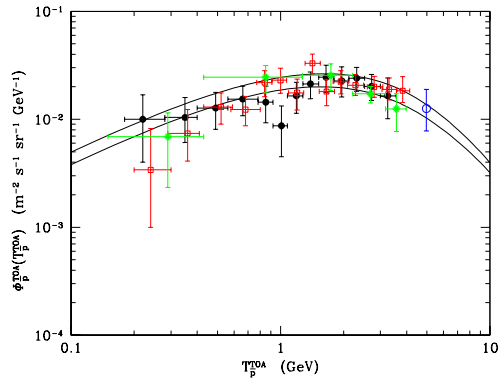}}
\caption{
The secondary antiproton flux lies within the uncertainty band -- delineated by
the two solid black lines -- which corresponds to all the possible propagation
schemes that are compatible with the B/C data. Experimental
observations at solar minimum are featured, 
in good agreement with the theoretical estimate. 
Full circles: {\sc bess} 1995-97 \cite{bess95-97}; 
open squares {\sc bess} 1998 \cite{bess98}; 
stars:  {\sc ams} \cite{ams98};
empty circles: {\sc caprice} \cite{caprice}.
}
\label{fig:sec_band}
\end{figure}

As a matter of fact, the conventional background for antiprotons is
produced in the galactic disk by the spallation of proton and helium
cosmic rays on the interstellar medium. A calculation of the flux at
the Earth of that population of secondary antiprotons has been performed
in Ref.~\cite{pbar_sec} with the same propagation model as for the B/C
analysis. The secondary flux has been derived consistently by employing
the propagation parameters that have been obtained from the full and
systematic analysis of Ref.~\cite{PaperI} on stable nuclei. 
The interstellar (IS) fluxes have been solar-modulated according to the force
field approximation in order to obtain top--of--atmosphere (TOA) fluxes. 
Throughout this paper, we compare our results with data taken during minimal
solar activity periods and we fix the modulation potential $\phi$ to 500 MV. 
It was found that the agreement between low-energy antiproton data and the 
estimate is excellent. Furthermore, the band within which the secondary flux
lies is fairly restrained as shown in Fig.~\ref{fig:sec_band}. In
Ref.~\cite{pbar_sec} it was actually
inferred a $\sim$ 20\% astrophysical uncertainty -- related to the
propagation model -- of the same order of magnitude as the theoretical
error arising from nuclear physics.
%
%
Taking adiabatic energy losses into account results -- below a kinetic energy
of 1 GeV -- into a small increase of our new secondary flux with respect to
the previously published spectrum of Ref.~\cite{pbar_sec}.

Constraining cosmic ray propagation with B/C provides a strong handle on any
species that originates from the disk of the Milky Way and allows in particular
the astrophysical uncertainties on secondary antiprotons to be mild as discussed
above. At variance with that secondary component, primary antiprotons are produced
by the annihilations of neutralinos spread all over the galactic halo. As
extensively discussed in Ref.~\cite{pbar_susy}, the corresponding uncertainties
span now two orders of magnitude, when the propagation parameters are varied between
the minimal and maximal configurations presented in Tab.~\ref{table:prop}.
%
\begin{table}[h!]
\begin{center}
{\begin{tabular}{@{}|c|c|c|c|c|c|@{}}
\hline
{\rm case} &  $\delta$  & $K_0$                 & $L$   & $V_{c}$       & $V_{A}$       \\
           &            & [${\rm kpc^{2}/Myr}$] & [kpc] & [km s$^{-1}$] & [km s$^{-1}$] \\
\hline
\hline
{\rm max} &  0.46  & 0.0765 & 15 & 5    & 117.6  \\
{\rm med} &  0.70  & 0.0112 & 4  & 12   &  52.9  \\
{\rm min} &  0.85  & 0.0016 & 1  & 13.5 &  22.4  \\
\hline
\end{tabular}}
\caption{
Astrophysical parameters compatible with B/C analysis and yielding
the maximal, median and minimal primary antiproton flux.
\label{table:prop}}
\end{center}
\end{table}
%
Primary fluxes depend sensitively on the thickness of the confinement
layers and also on the convective wind that wipes cosmic ray species
away from the galactic disk. It is not surprising therefore if the
largest primary antiproton yield corresponds to the combination
$L = 15$ kpc and $V_{c} = 5$ km s$^{-1}$, whereas the smallest flux
obtains when $L = 1$ kpc and $V_{c} = 13.5$ km s$^{-1}$.
The Alfv\'enic velocity $V_{A}$ is strongly correlated with the normalization
constant $K_{0}$ because the B/C ratio determines the diffusive reacceleration
parameter $K_{EE} \propto V_{A}^{2} / K_{0}$ -- see Ref.~\cite{pbar_susy} --
with an accuracy of $\pm 15$\%. The  parameter $V_{A}$ will not be further discussed.
Finally, we remind that -- as obtained in Refs.~\cite{pbar_susy,lowind} -- the
specific form assumed for the dark matter distribution in the Galaxy is fairly
irrelevant in the calculation of the propagated primary antiproton flux.
In particular, for not too thick confinement layers and strong convection winds,
the cosmic ray diffusion range is small and solar circle abundances are blind
to the center of the Milky Way and its putative neutralino cusp.

\section{Results and conclusions}
\label{sect:results}

%
\begin{figure*}[t]
\vskip -2.0cm
{\includegraphics[width=\columnwidth]{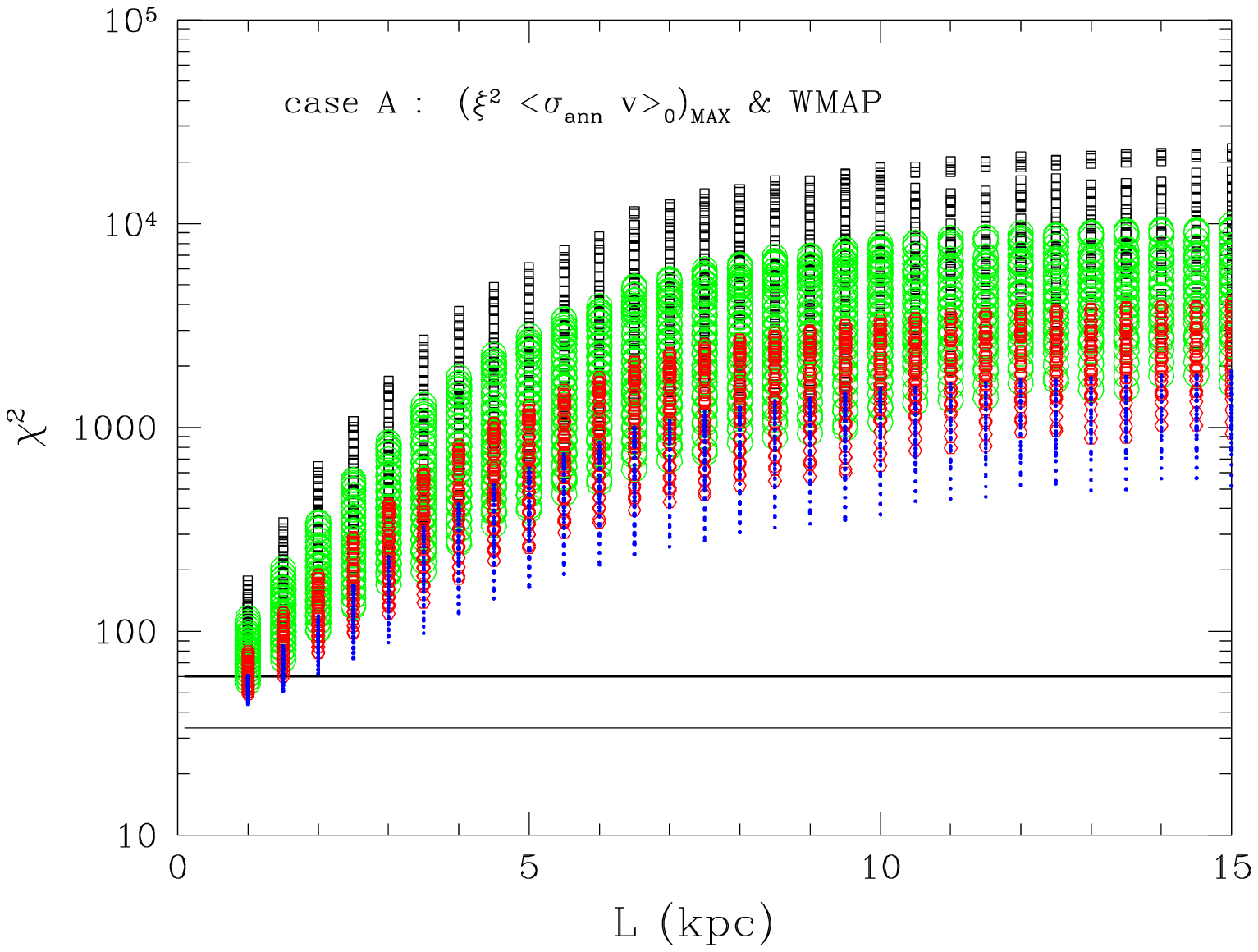}}
{\includegraphics[width=\columnwidth]{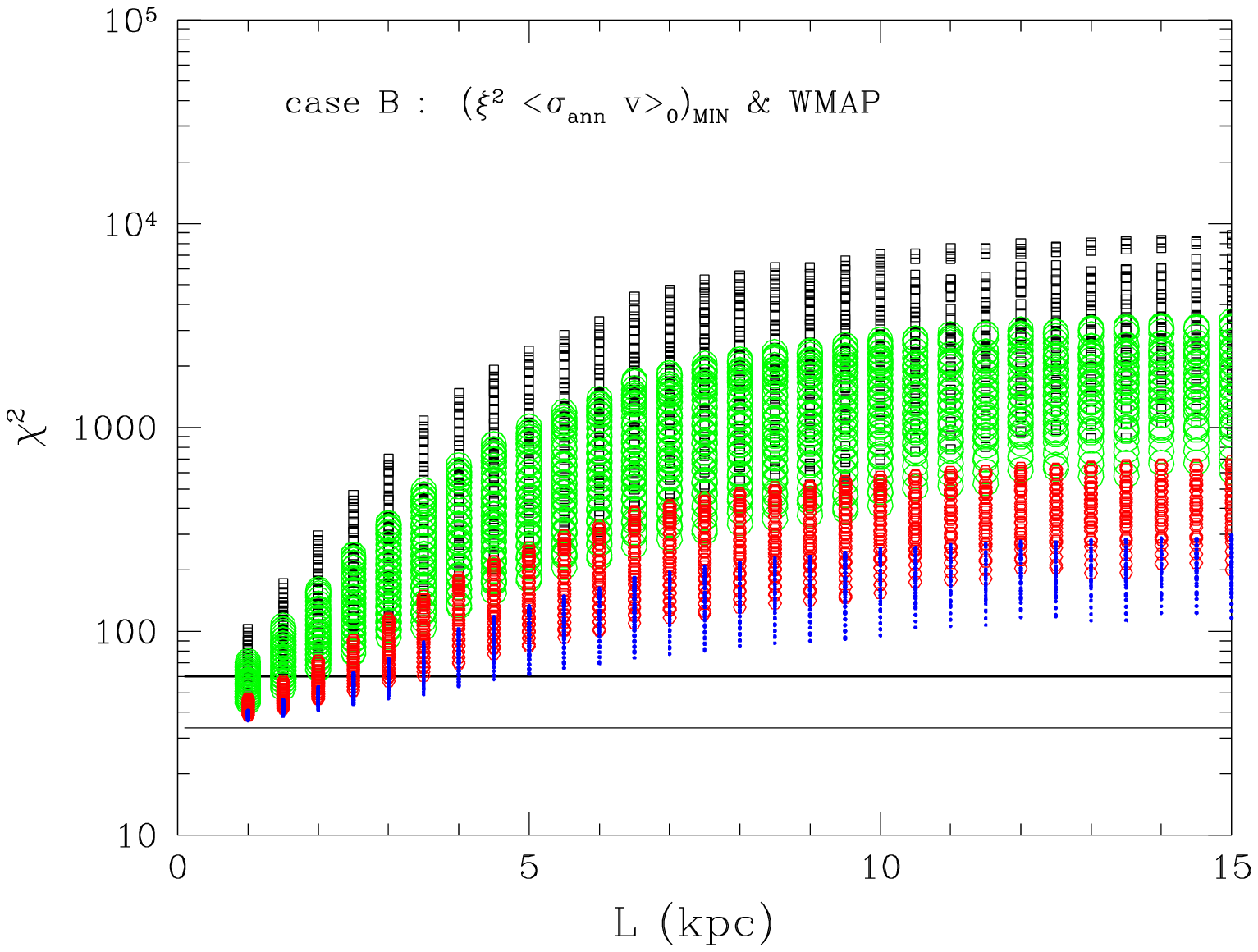}}
\vskip -2.0cm
{\includegraphics[width=\columnwidth]{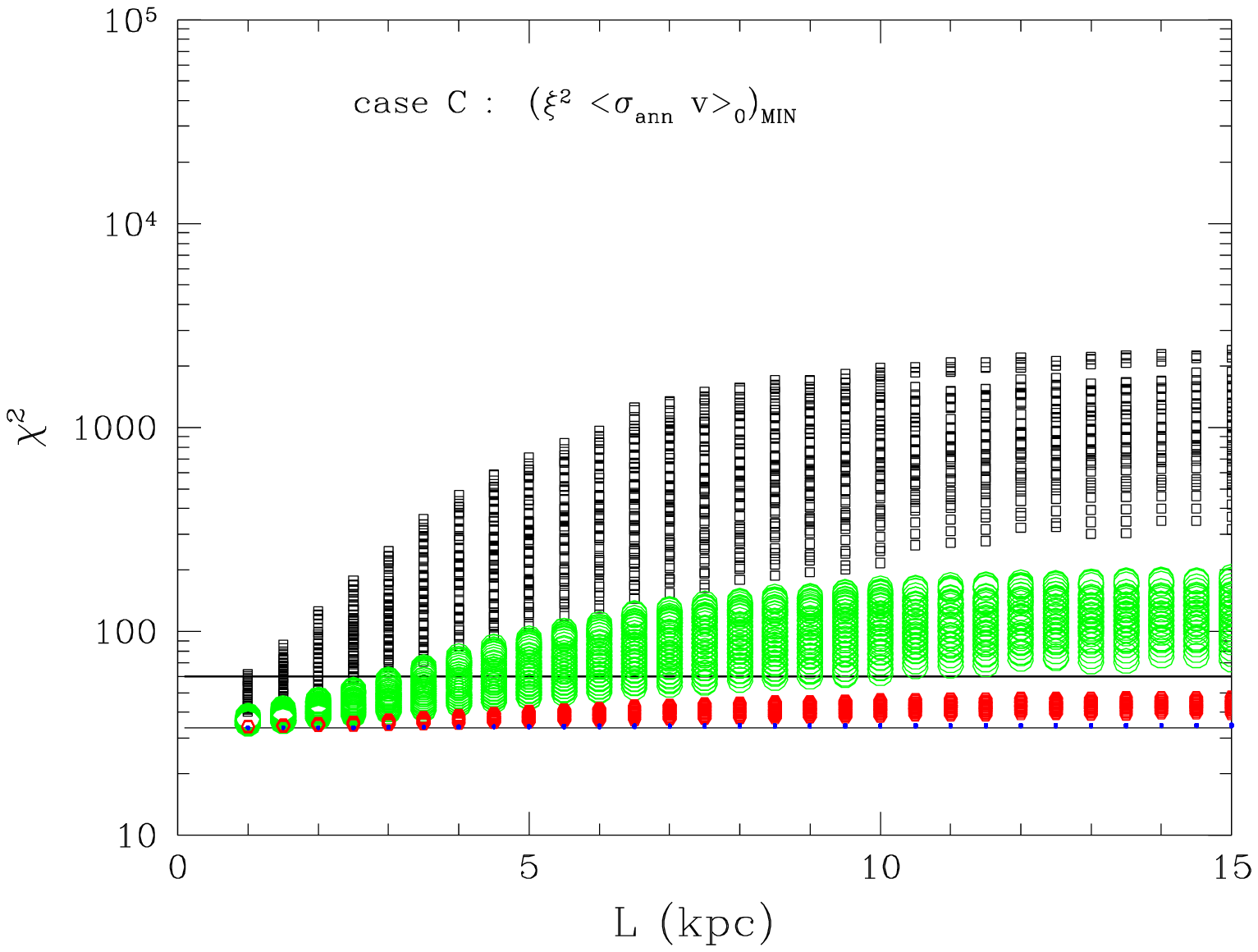}}
\vskip -1.0cm
\caption{
The $\chi^{2}$ is featured as a function of the diffusive halo thickness $L$.
Black squares, green big circles, red smaller circles and blue dots
correspond to a neutralino mass $m_{\chi}$ of 10, 20, 30 and 40 GeV, respectively. 
At fixed
neutralino mass $m_{\chi}$, each point is associated to a specific combination
of the galactic propagation parameters yielding a good fit for the B/C ratio.
Panels a, b and c respectively correspond to cases A), B) and C) defined 
in  the text.
The upper horizontal line indicates the discriminating $\chi^{2}$ value
above which the fit to the low-energy antiproton data is no longer acceptable.
The lower horizontal line corresponds to the $\chi^{2}$ calculated with the
secondary component alone for the median configuration of Tab.~\ref{table:prop}.
}
\label{fig:chi2_L}
\end{figure*}
%

Because our analysis is focused on light neutralinos
-- $m_{\chi} \leq 40$ GeV -- our main interest is
in low-energy antiprotons. That is why we have selected the observations
performed in the GeV region by
Bess 1995-97 \cite{bess95-97},
Bess 1998    \cite{bess98} and
AMS 1998     \cite{ams98}.
All these experiments operated at solar minimum for which $\phi = 500$ MV.
The corresponding 32 data points are presented in Fig.~\ref{fig:sec_band}
and are already very well explained by a pure secondary component. Actually,
we have derived a $\chi^{2}$ of 33.6 in the case of the median configuration
of Tab.~\ref{table:prop}. As discussed in Sec.~\ref{subsect:cr_propagation},
should the galactic cosmic ray propagation parameters be varied so as to keep
the B/C reduced $\chi^{2}$ below a generous value of 1.8, the secondary
antiproton flux would change by less than 20\% -- see Ref.~\cite{pbar_sec}.
To put our discussion on a quantitative basis, we need to establish a
criterion for deciding whether or not a specific supersymmetric configuration
is excluded by the above mentioned antiproton data. To this purpose, we compute
the primary flux which such a configuration yields, add it to the secondary
component and derive the corresponding $\chi^{2}$. Notice that Caprice
\cite{caprice} has been disregarded in our calculation as it operated
at a slightly higher energy than the range in which we are interested.
Then we decide that a specific supersymmetric configuration is excluded if
the corresponding $\chi^{2}$ exceeds a critical value of 60. For 32 degrees of
freedom, this translates into a disagreement at the 99.5 \% C.L.
It is worth stressing that our exclusion criterion is purposely taken
on the conservative side.

A final word of caution is in order before we proceed. The problem is complicated
by the fact that the galactic cosmic ray propagation model is not unique.
Actually, in the present analysis, we have varied the five propagation
parameters presented in Sec.~\ref{subsect:cr_propagation} and considered
all the possible combinations that are in good agreement with stable nuclei
\cite{PaperI}. We keep the conservative attitude of selecting propagation
models as long as they lead to a reduced $\chi^{2}_{\rm red \, B/C} \leq 1.8$
on the B/C data.

To commence, we focus our investigation on the supersymmetric configurations
which the present data on low-energy cosmic ray antiprotons constrain most.
Those configurations are associated to large antiproton yields and
correspond to high values of the effective annihilation cross-section
$\xi^{2} <\sigma_{\rm ann} v>_{0}$. We concentrate here on cases A) and B)
for which the neutralino relic density is relevant to cosmology.
Case A) corresponds to the largest antiproton signal and consequently to the
strongest constraint as is clear in the panel a of Fig.~\ref{fig:chi2_L}
where the $\chi^{2}$ is plotted as a function of the diffusive halo
thickness $L$. (Black) squares, (green) big circles, (red) smaller circles and
(blue) dots correspond to a neutralino mass $m_{\chi}$ of 10,
20, 30 and 40 GeV, respectively. The constellation of points of the same 
shape (and color) -- which
corresponds to a specific neutralino mass -- is obtained by scanning the
entire set of cosmic ray propagation models that are compatible with B/C
-- as discussed above. The effective annihilation cross-section
$\xi^{2} <\sigma_{\rm ann} v>_{0}$ reaches here a maximal value of
$10^{-36}$ cm$^{2}$ as may be readily inferred from the scatter plot of
Fig.~\ref{fig:xi2_sigma0} and its upper boundary.
Case B) corresponds to the minimal value of the annihilation
cross-section $<\sigma_{\rm ann} v>_{0}$ that is yet compatible with WMAP.
The neutralino relic abundance is set to the upper bound
$(\Omega_{\rm CDM})_{\rm max} h^2 = 0.13$. The corresponding scatter plot
is presented in panel b of Fig.~\ref{fig:chi2_L}.
The lower horizontal line at $\chi^{2} = 33.6$ corresponds to the fit
with the secondary antiproton component alone. The upper horizontal line
indicates the discriminating $\chi^{2}$ value above which the fit to the
low-energy antiproton data is no longer acceptable.
The large spread exhibited by the scatter plots of Fig.~\ref{fig:chi2_L}
illustrates the strong sensitivity of the primary antiproton flux to
the astrophysical parameters that account for cosmic ray propagation
throughout the Milky Way and its halo. This is at variance with the
stability of the secondary flux as already remarked above.
As is clear from panels a and b, light neutralinos of cosmological interest
-- {\it {i.e.}} with a relic abundance in the WMAP range -- are compatible
with the present data on cosmic ray antiprotons only if the parameter $L$ is
on the very low side of its physical interval. Should the diffusive halo
thickness $L$ turn out to be larger than 2 kpc -- a quite realistic
assumption given the existence of electron synchrotron radiation far
above the galactic disk -- neutralinos lighter than 20 GeV would be
excluded. If we now assume that $L \geq 5$ kpc, we readily conclude
that neutralinos with sizeable relic abundance must be heavier than
40 GeV.
%
\begin{figure}[h!]
\vskip -2.0cm
{\includegraphics[width=\columnwidth]{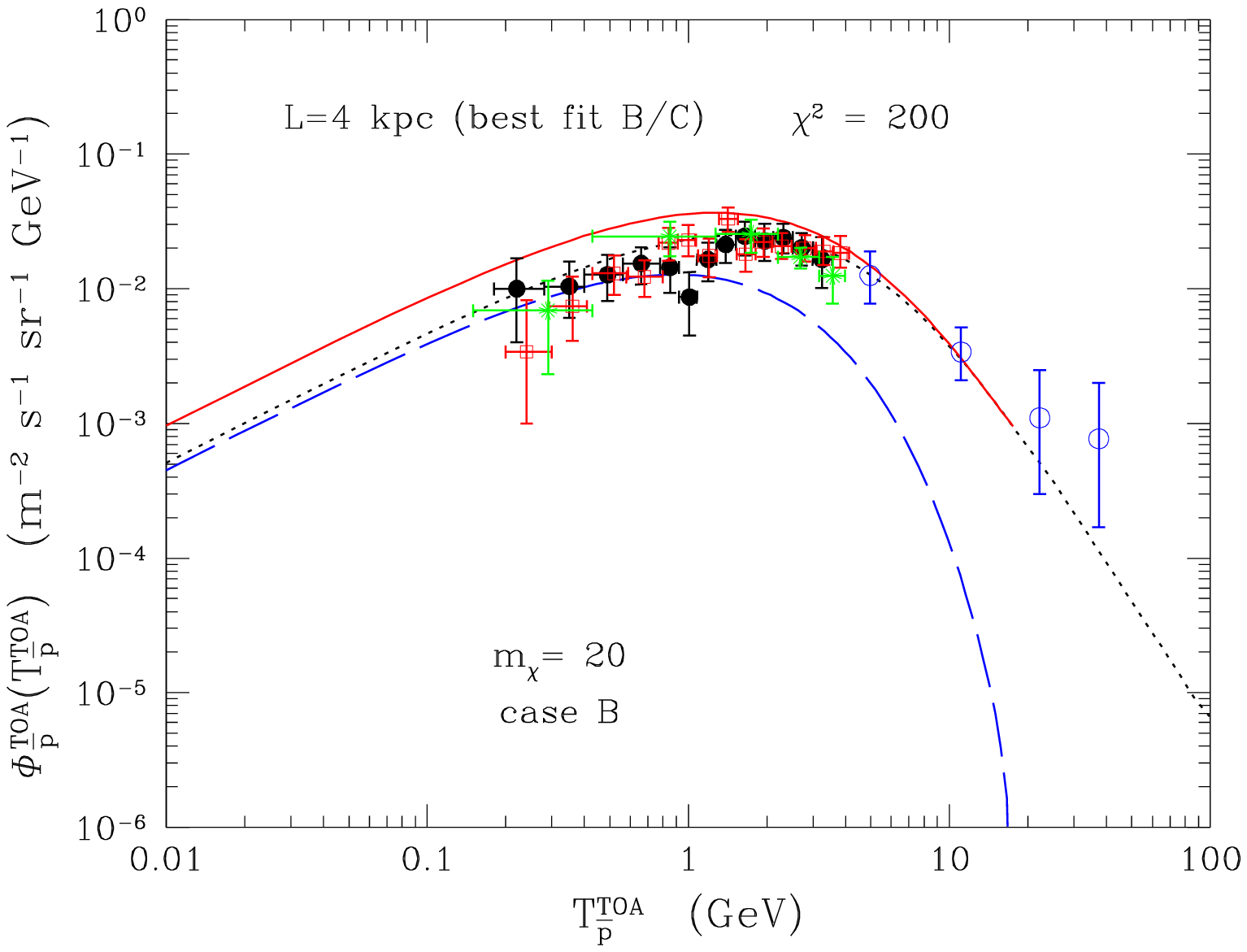}}
\vskip -1.0cm
\caption{
The primary antiproton spectrum -- blue long-dashed curve -- has been computed
for a 20 GeV neutralino and corresponds to the case B) of a maximal relic
abundance. The median configuration of Tab.~\ref{table:prop} has been assumed.
When added to the secondary component -- blue dotted line -- the primary signal
results into a total antiproton flux -- red solid spectrum -- well above the data
points. We actually infer a $\chi^{2}$ value of 200 to be compared to the
exclusion limit of 60. 
Symbols for experimental data as in Fig. \ref{fig:sec_band}.
}
\label{fig:spectra_m20_case_b_best_fit_BC}
\end{figure}
%
In order to illustrate how our $\chi^{2}$ exclusion criterion
translates on the observations, we have plotted in
Fig.~\ref{fig:spectra_m20_case_b_best_fit_BC} the total antiproton spectrum
-- (red) solid curve -- for a 20 GeV neutralino with maximal relic abundance.
The median configuration of Tab.~\ref{table:prop} has been selected for the
cosmic ray propagation parameters. It corresponds to the best fit to the B/C data.
The primary (blue long-dashed curve) and secondary (blue dotted line)
components add up to yield an antiproton signal well in excess of the observations.
The corresponding $\chi^{2}$ reaches actually the unacceptable value of 200.

%
\begin{figure}[h!]
\vskip -2.0cm
{\includegraphics[width=\columnwidth]{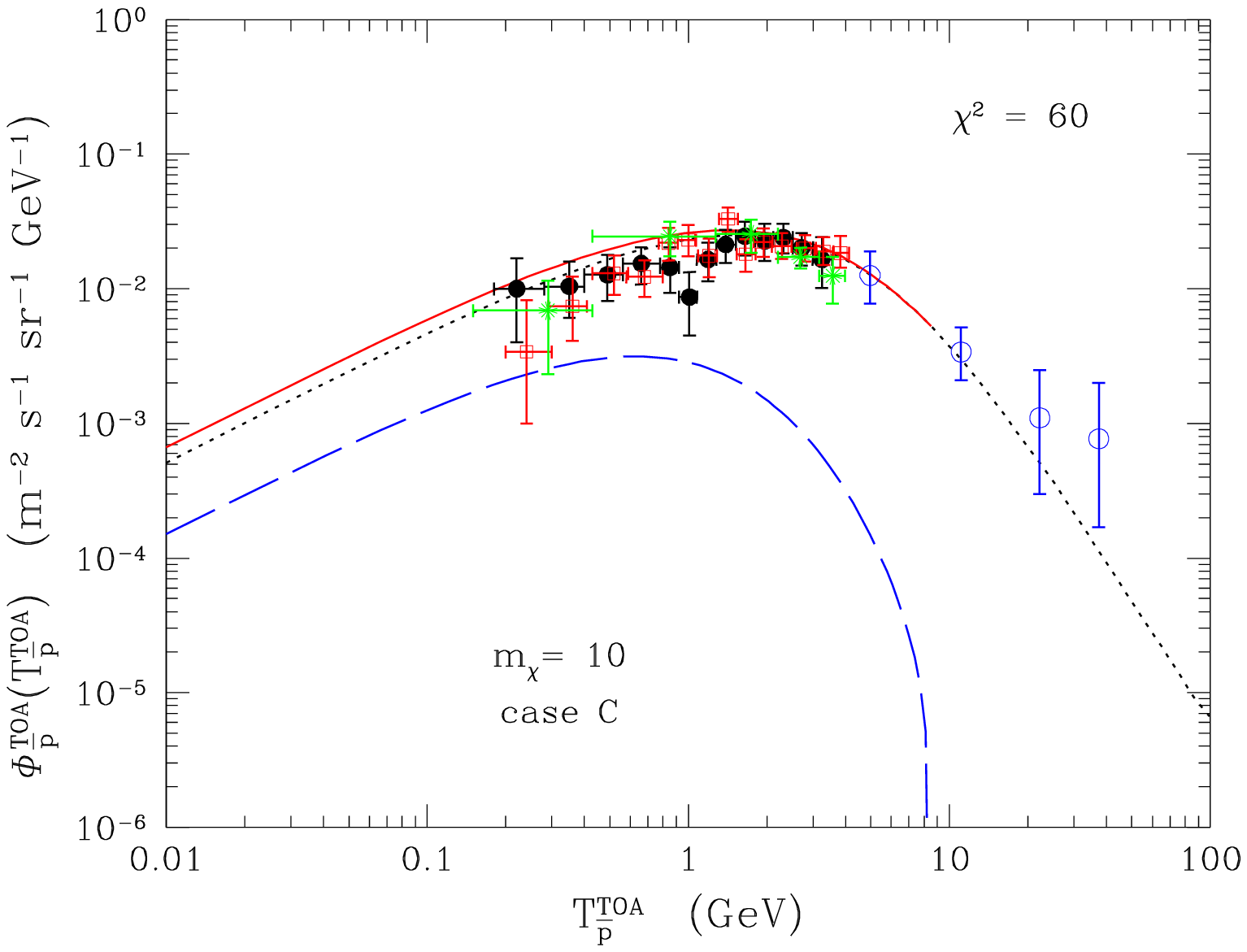}}
\vskip -1.0cm
\caption{
The antiproton spectrum is featured together with its primary -- blue
long-dashed curve -- and secondary -- blue dotted line -- components
for a 10 GeV neutralino. The supersymmetric configuration corresponds
to maximal rescaling -- case C) -- and the specific set of astrophysical
parameters that has been extracted from panel c of Fig.~\ref{fig:chi2_L}.
This case corresponds to $\chi^2$=60, and then 
is marginally acceptable. 
Symbols for experimental data as in Fig. \ref{fig:sec_band}.
}
\label{fig:spectra_m10_case_c}
\end{figure}
%

%
\begin{figure}[h!]
\vskip -2.0cm
{\includegraphics[width=\columnwidth]{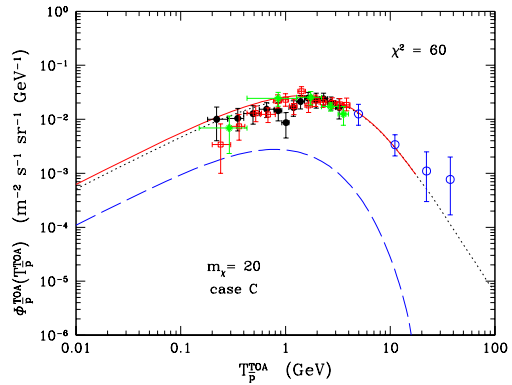}}
\vskip -1.0cm
\caption{
Same as in  Fig.\ref{fig:spectra_m10_case_c} with a 20 GeV neutralino.
}
\label{fig:spectra_m20_case_c}
\end{figure}
%

The last panel of Fig.~\ref{fig:chi2_L} features supersymmetric configurations
for which the effective annihilation cross-section $\xi^{2} <\sigma_{\rm ann} v>_{0}$
is minimal and rescaling is maximal. The antiproton signal is at its weakest
level and we do not expect low-energy antiproton
data to be very constraining. Panel c
of Fig.~\ref{fig:chi2_L} indicates nevertheless that a 10 GeV neutralino is
excluded provided that the diffusive halo thickness $L$ exceeds 2.5 kpc.
Should $L$ be larger than 10 kpc, the limit rises to 20 GeV. That panel shows
also to what extent cosmologically subdominant neutralinos with a mass in the
range 20 GeV $\lsim m_{\chi} \lsim $ 40 GeV escape conflict with present data.
A potential distortion is difficult to unravel from the antiproton spectrum
given the available observations because error bars are too large. Actually,
Fig.~\ref{fig:spectra_m10_case_c} and \ref{fig:spectra_m20_case_c} feature
respectively the primary antiproton flux of a 10 and 20 GeV neutralino 
(blue long-dashed curves) in the case C) of maximal rescaling together with
the conventional secondary component (blue dotted lines). The cosmic ray
propagation configurations that have been extracted from the panel c of
Fig.~\ref{fig:chi2_L} -- a (black) square if the neutralino mass is 10 GeV
or a (green) big circle when $m_{\chi}$ is twice as large -- lie on the upper
horizontal line for which the fit to the low-energy antiproton data yields
a $\chi^{2}$ value of 60. Below an antiproton kinetic energy of $\sim$ 1 GeV,
the global signal represented by the (red) solid curves is slightly shifted
upwards with respect to the secondary spectrum whereas above 1 GeV, both
spectra are identical.
%
\begin{figure*}[t!]
\vskip -2.0cm
{\includegraphics[width=\columnwidth]{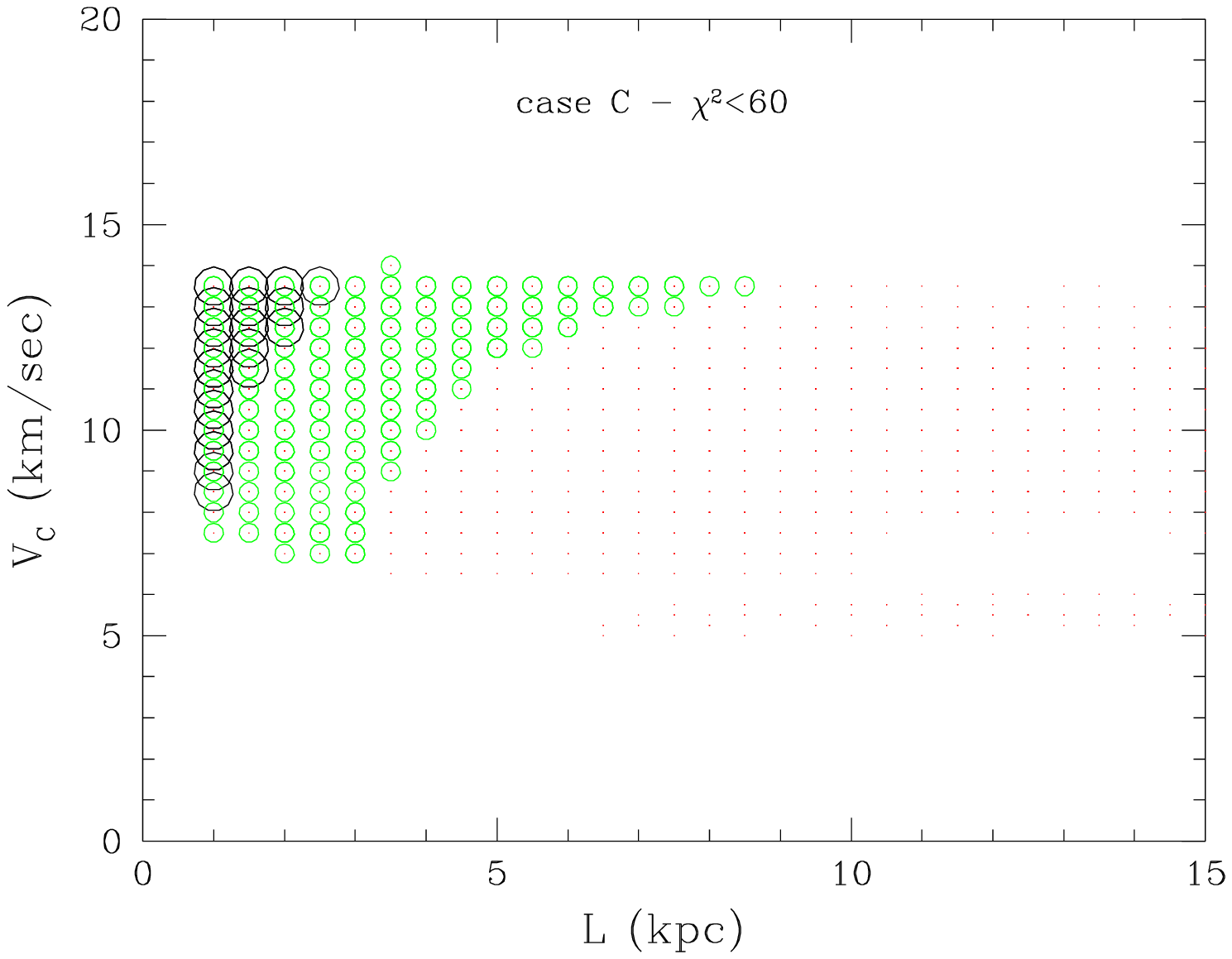}}
{\includegraphics[width=\columnwidth]{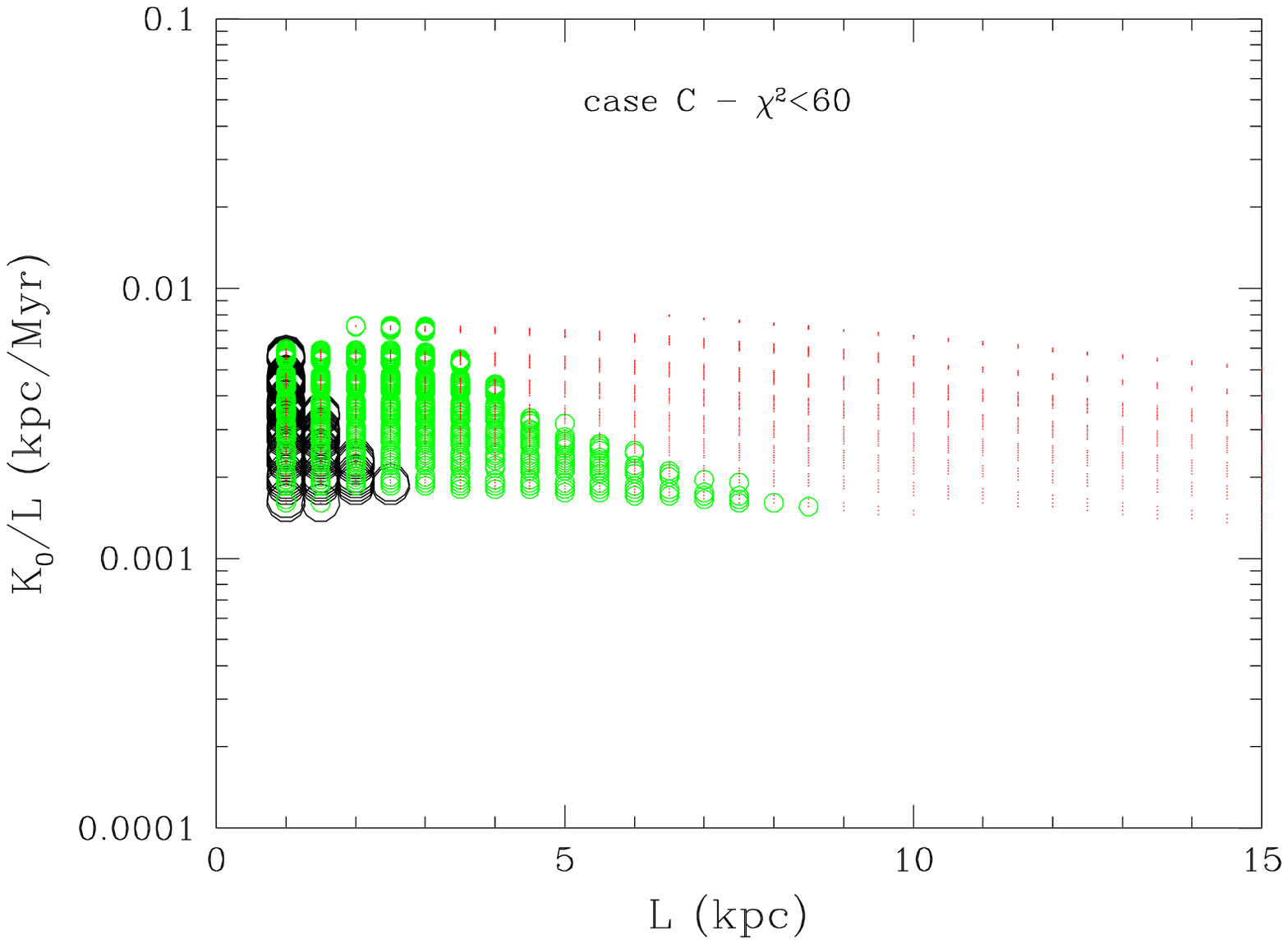}}
\vskip -2.0cm
{\includegraphics[width=\columnwidth]{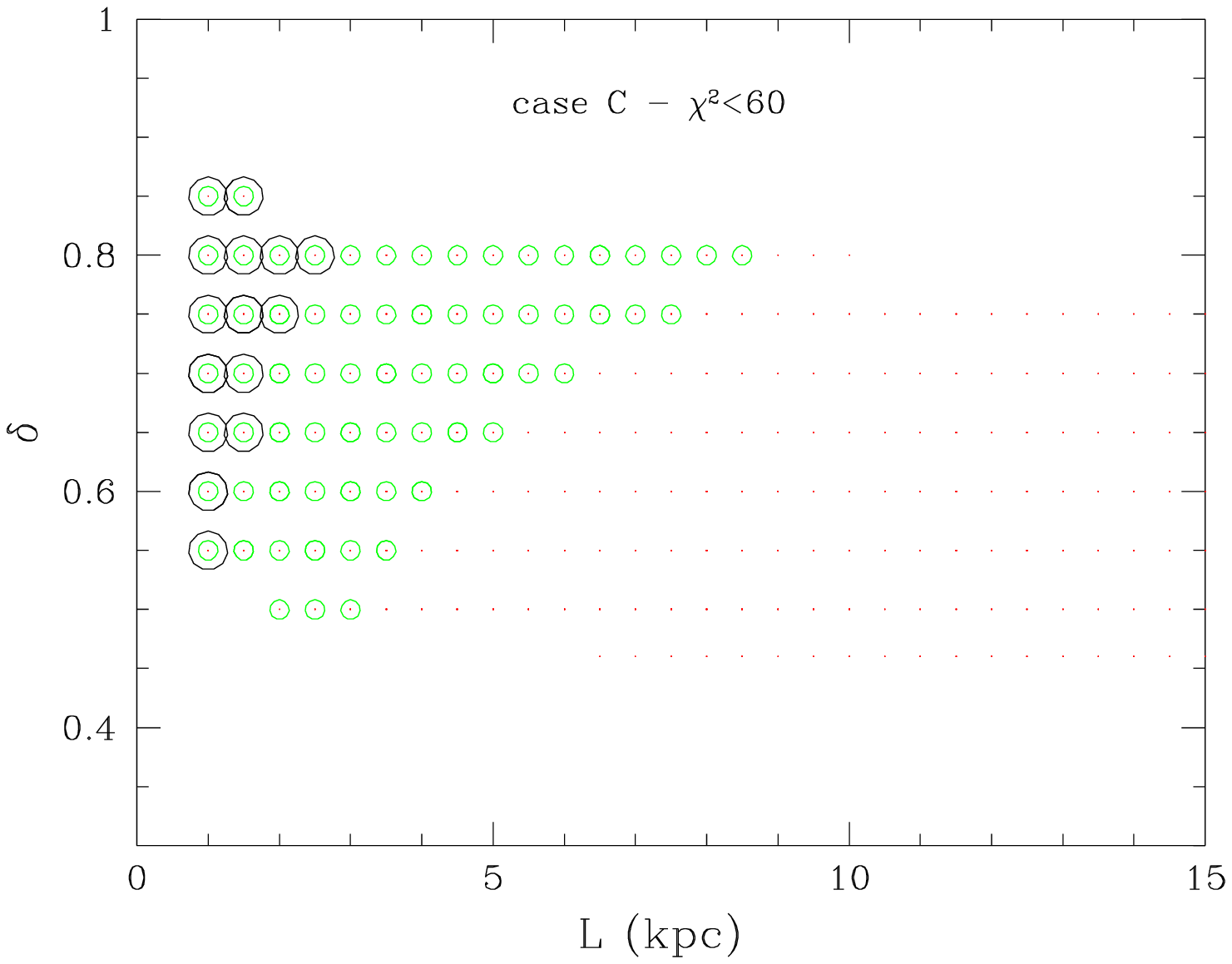}}
\vskip -1.0cm
\caption{
The three panels feature the same selection of astrophysical
configurations from panel c of Fig.~\ref{fig:chi2_L} for which
the fit to the antiproton data is acceptable with $\chi^{2} \leq 60$.
Big black circles, smaller green circles and red dots respectively
correspond to a neutralino mass $m_{\chi}$ of 10, 20 and 30 GeV.
The panels successively display the cosmic ray propagation parameters
$V_{c}$ , $K_{0} / L$ and $\delta$ as a function of the diffusive halo
thickness $L$.
}
\label{fig:case_c_propagation_OK}
\end{figure*}
%
Richer samples of experimental data are a necessary condition for disentangling
a supersymmetric signal from background secondary antiprotons. The spectrum
of the latter suffers nevertheless from theoretical uncertainties -- cosmic
ray propagation and nuclear cross-sections -- that need to be further reduced
in order to make antiproton measurements a useful probe for cosmologically
subdominant neutralinos in the 10 to 30 GeV range.

It is very instructive to pursue the analysis of the maximal rescaling case C)
in terms of other astrophysical parameters. The three scatter plots of
Fig.~\ref{fig:case_c_propagation_OK} feature the same selection of astrophysical
configurations drawn from Fig.~\ref{fig:chi2_L} -- panel c -- for which
the fit to the antiproton data is acceptable with $\chi^{2} \leq 60$. Big
(black) circles, smaller (green) circles and (red) dots  correspond
to a neutralino mass $m_{\chi}$ of 10, 20 and 30 GeV respectively. The smaller the latter,
the more abundant the neutralinos at fixed mass density and the stronger
the antiproton annihilation signal. This trend is clear in each of the panels
of Fig.~\ref{fig:chi2_L}. It is even clearer in panel c. The relevant
annihilation cross-section is in that case the minimal value of
$\xi^{2} <\sigma_{\rm ann} v>_{0}$ which increases rapidly with decreasing neutralino
mass -- see the lower boundary of the scatter plot in Fig.~\ref{fig:xi2_sigma0}.
Low $m_{\chi}$ configurations are allowed if the primary antiproton excess
which they yield does not propagate efficiently from the halo of
the Milky Way to the solar system. Thin diffusive halos and strong galactic winds
are preferred. This is particularly obvious for a 10 GeV neutralino. Notice how
the big (black) circles concentrate in the upper-left corner of the $V_{c}$ versus
$L$ diagram. Remark also how the normalization $K_{0}$ of the diffusion coefficient
scales with the diffusive halo thickness $L$. In the second panel of
Fig.~\ref{fig:case_c_propagation_OK}, the ratio $K_{0} / L$ lies actually in the
range between $10^{-3}$ and $10^{-2}$ kpc Myr$^{-1}$ irrespective of $L$. That
ratio turns out to be crucial in the diffusion equation. Finally, we do not find
any particular correlation between $\delta$ and $L$ even if small values for the
latter are still pointed towards in the last panel of
Fig.~\ref{fig:case_c_propagation_OK}.

%
\begin{figure}[h!]
\vskip -2.0cm
{\includegraphics[width=\columnwidth]{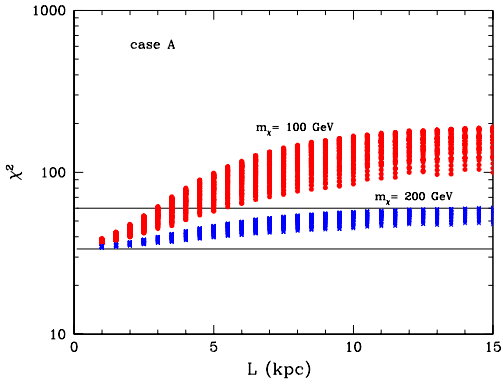}}
\vskip -1.0cm
\caption{
Same as in Fig.~\ref{fig:chi2_L}, panel a) but for $m_{\chi}$=100 GeV
-- upper branch -- and $m_{\chi}$=200 GeV -- lower branch. Horizontal
lines are the same as in Fig.~\ref{fig:chi2_L}.
}
\label{fig:chi2_L_m100_m200}
\end{figure}
%

%
\begin{figure}[h!]
\vskip -2.0cm
{\includegraphics[width=\columnwidth]{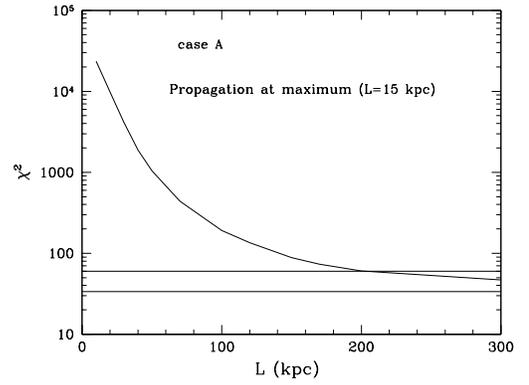}}
\vskip -1.0cm
\caption{
The $\chi^{2}$ is plotted as a function of the neutralino mass. 
Case A) has been
selected with astrophysical parameters leading to the maximal signal. Horizontal
lines are the same as in Fig.~\ref{fig:chi2_L}.
}
\label{fig:chi2_mchi_case_a_astromax}
\end{figure}
%

To conclude our investigation, we would like to assess the potential of
low-energy antiproton measurements for discovering heavier neutralinos.
For that purpose, we present in the $\chi^{2}$ versus $L$ plane of
Fig.~\ref{fig:chi2_L_m100_m200} the same kind of scatter plot as in panel a
of Fig.~\ref{fig:chi2_L}. Neutralino masses of 100 GeV (red stars) and 200
GeV (blue crosses) have been displayed. In the supersymmetric configuration
which we have selected for each neutralino mass, the effective annihilation
cross-section $\xi^{2} <\sigma_{\rm ann} v>_{0}$ reaches its maximal value
-- case A) -- so that the antiproton signal is the strongest. At fixed
$m_{\chi}$, each star or cross is associated to a combination of galactic
propagation parameters yielding a good fit for the B/C ratio. Notice how
the constellation of (blue) crosses lies between the two horizontal lines.
We readily conclude that whatever the cosmic ray propagation model is,
low-energy antiproton data are useless as long as $m_{\chi} \gsim$ 200 GeV.
Our statement is supported by Fig.~\ref{fig:chi2_mchi_case_a_astromax} where
the maximal possible value of $\chi^{2}$ has been plotted as a function of
neutralino mass. The effective annihilation cross-section $\xi^{2} <\sigma_{\rm
ann} v>_{0}$ has been once again tuned to its maximal value -- case A) -- and
the astrophysical configuration max of Tab.~\ref{table:prop} has been chosen.
The $\chi^{2}$ curve drops below the exclusion value of 60 when the neutralino
mass exceeds 200 GeV.
\\
Observations at significantly higher energies will probably be necessary to
explore the regime where neutralinos are heavy. We leave such an investigation
for a future publication but we cannot resist noticing that the Caprice data
above 10 GeV already exhibit an excess with respect to the secondary antiproton
background. If such a distortion is confirmed -- by the forthcoming
Pamela satellite mission \cite{pamela} for instance -- it should have
to be explained \cite{pbar_susy}. Pamela will actually probe the antiproton
spectrum between 80 MeV and 190 GeV and collect data during three years
starting at the end of 2005. 
As well as the Pamela experiment, other space missions are planned  - such as
AMS-02 on board the International Spatial Station \cite{ams02} or the
balloon-borne Bess Polar mission in Antarctica \cite{bess_polar} - or in
project - such as the GAPS satellite experiment \cite{gaps}. They will 
measure the low-energy antiproton spectrum with very high accuracy, and their
results will be of great relevance for improving the study developed in the
present paper. 
Improvements in the understanding of the propagation of galactic cosmic rays are
foreseeable in the next future, thanks to long and ultra-long duration balloon
missions \cite{cream} and space-based expriments \cite{ams02}. 
Even if the best expectations are deserved to the determination of the diffusion
coefficient power spectrum $\delta$, data on the B/C quantity in the GeV/n
region will also constrain the diffusive halo thickness $L$ 
\cite{wwp} which, among the
astrophysical parameters, plays the most important role in our previous
analysis. 

Finally, we recall that the conclusions of the present paper are drawn
for the case of a smooth distribution of dark matter in the galactic halo.
Should the dark matter distribution have some clumpiness, the antiproton
signal would be enhanced \cite{clump}, 
and consequently some of the constraints on the previously
discussed supersymmetric configurations would become more stringent.

\newpage

\acknowledgments
We acknowledge Research Grants funded jointly by the
Italian Ministero dell'Istruzione, dell'Universit\`a e della Ricerca
(MIUR), by the University of Torino and by the Istituto Nazionale di
Fisica Nucleare (INFN) within the {\sl Astroparticle Physics Project}.
P.S. acknowledges a support from the French Programme National de
Cosmologie PNC.
F.D. acknowledges support from the A. von Humboldt Stiftung and hospitality
from the  Max-Planck Institute f\"ur Physik in Munich, where part of this work
was done.


\end{document}